\documentstyle[osa,12pt]{revtex}

\begin{document}
\title{High Temperature Thermopower in $La_{2/3}Ca_{1/3}MnO_3$ Films:\\
Evidence for Polaronic Transport}
\author{M. Jaime and M. B. Salamon}
\address{Departament of Physics, University of Illinois at Urbana-Champaign\\
1110 W. Green Street, Urbana IL\ 61801-3080}
\author{M. Rubinstein, R. E. Treece, J. S. Horwitz and D. B. Chrisey}
\address{U. S. Naval Research Laboratory, Washington D.C. 20375-5000}
\date{7/19/96}
\maketitle

\begin{abstract}
Thermoelectric power, electrical resistivity and magnetization experiments,
performed in the paramagnetic phase of $La_{2/3}Ca_{1/3}MnO_3,$ provide
evidence for polaron-dominated conduction in CMR materials. At high
temperatures, a large, nearly field-independent difference between the
activation energies for resistivity $\rho $ and thermopower $S$, a
characteristic of {\em Holstein Polarons,} is observed, and $\ln \rho $
ceases to scale with the magnetization$.$ On approaching T$_C$, both
energies become field-dependent, indicating that the polarons are
magnetically polarized. Below T$_C$, the thermopower follows a law $S(H)\sim
1/\rho (H)$ as in non saturated ferromagnetic metals.

\end{abstract}

\newpage \ Since Chahara et al.\cite{Chahara,vonHel2} successfully prepared
films showing ferromagnetism and novel magnetoresistance properties,
perovskite-based materials of composition $La_{1-x}A_xMnO_3$ ($A=Sr,Ca,Ba$)
have received wide attention from theoretical and experimental researchers,
with the consequent proliferation of new models and a rapid improvement in
sample quality. A broadening of the magnetization curves, accompanied by a
reduction of several orders of magnitude in the electrical resistance,
occurs when a magnetic field of the order of $5$T is applied. The effect has
been called colossal magnetoresistance ({\em CMR}) and attributed to the
coherent hopping of electrons between spin-aligned $Mn^{3+}$ and $Mn^{4+}$
ions. In spite of the effort expended, there is still no consensus as to the
microscopic mechanisms causing {\em CMR}.

Reports of large, negative magnetoresistance ({\em MR})\ date from 1954 and,
because of the potential technological consequences, attracted considerable
attention over several decades.\cite{Volger,vonMol2} The magnetic and
transport properties of {\em CMR} materials were initially attributed solely
to the double exchange mechanism ({\em DE}) proposed by Zener.\cite{Zener} A
theoretical consideration of $La_{1-x}Sr_xMnO_3$ predicted \cite{Mazaferro}
band broadening at the ferromagnetic transition, reducing the gap and
leading to increased or metallic conductivity in the {\em FM} state and
activated conductivity above $T_C$. However, magnetic interactions alone are
insufficient to explain the observed {\em CMR}.\cite{Millis} Recently, we
reported low temperature thermopower ($S$) and resistivity ($\rho $) data in
partially annealed $La_{2/3}Ca_{1/3}MnO_3$ films suggesting that transport
via small lattice polarons could be important above $T_C$ .\cite{Jaime} In
this Letter we present high temperature results on well-annealed samples,
providing the first strong evidence for magnetic polaron transport in $%
La_{2/3}Ca_{1/3}MnO_3$.

The formation and transport properties of small lattice polarons in{\em \ }%
strong electron-phonon ({\em e-ph}) coupled systems, in which charge
carriers are susceptible to self-localization in energetically favorable
lattice distortions, was first discussed in disordered materials \cite
{Holstein} and later extended to crystals.\cite{MottDavis,Emin,Moliton} In a
parallel development, Kasuya and Yanase \cite{Kasuya} considered the
behavior of purely magnetic polarons ({\em MP}), defined as a carrier
localized at impurity centers by a polarization cloud, the transport
mechanism being thermal hopping between sites. In this picture, the hopping
activation energy disappears if the moments are aligned by an external
magnetic field; the material transforms from a semiconductor to a dirty
metal, exhibiting a large negative {\em MR}. On the other hand Mott\cite
{MottDavis} argued that the mobility of purely magnetic polarons is
diffusive in nature, {\em i.e.} has a power law rather than thermally
activated temperature dependence. Emin \cite{Emin} considered the nature of
lattice polarons in magnetic semiconductors. In this model, magnetic
polarons are carriers self-localized by lattice distortions but also dressed
with a magnetic cloud. A transition from large to small polaron occurs as
the ferromagnet disorders, successfully explaining the metal-insulator
transition observed experimentally in $EuO$.

If the carrier together with its associated crystalline distortion is
comparable in size to the cell parameter, the object is called a small, or 
{\em Holstein, }polaron ({\em HP}). Because a number of sites in the crystal
lattice can be energetically equivalent, a band of localized states can
form. These energy bands are extremely narrow, see Fig.1, and the carrier
mobility associated with them is predominant only at very low temperatures.
It is important to note that these are not extended states even at the
highest temperatures. Electrical conduction then occurs via either quantum
tunneling ({\em QT}) or thermal hopping of the {\em HP} among sites. Three
different temperature regimes can be distinguished. At very low
temperatures, where $k_BT\leq 10^{-4}$ $eV$, the only possible mechanism is 
{\em QT} between neighboring distortions. As the temperature is raised, but
for $T\leq \theta _D/2$, half the Debye temperature, phonon assisted hopping
dominates producing a conductivity $\sigma \propto \exp (-T^{-1/4});$ this
is not, however, associated with variable range hopping. At the highest
temperatures the dominant mechanism is thermally activated hopping of
carriers, with an activated mobility $\mu _P=[c(1-c)ea^2/\hbar
](T_o/T)^s\exp [-(W_H-J^{3-2s})/k_BT]$ where $a$ is the hopping distance; $J$%
, the transfer integral; $c$, the polaron concentration; and $W_H,$ one half
of the polaron formation energy $E_p$. In the non-adiabatic limit, we have $%
s=3/2$ and $k_BT_o=\left( \pi J^4/4W_H\right) ^{1/3}$ and, in the adiabatic
limit, $s=1$ and $k_BT_o=\hbar \omega _o$, where $\omega _o$ is the optical
phonon frequency.\cite{raffaelli} The criterion for non-adiabatic behavior
is that the experimental $k_BT_o\ll \hbar \omega _o.$ Using experimental
values for $\rho $, $E_\rho $, $S$ and cell parameter we find that $%
k_BT_o/\hbar \simeq 10^{14}\;s^{-1},$ comparable to optical phonon
frequencies, although it could be considered a marginal case. We will assume
the adiabatic limit to hold, in which case the electrical conductivity, $%
\sigma =eN\mu _P$, where $N$ is the equilibrium polaron number at a given
temperature, can be expressed as

\begin{equation}
\sigma =\frac{c(1-c)e^2T_o}{\hbar aT}\exp \left( -\frac{\epsilon _0+W_H-J}{%
k_BT}\right)  \label{eqn1}
\end{equation}

Because the carrier hops from one locally distorted site to another that has
been thermally activated, it carries only the entropy associated with its
chemical potential, leading to a simple expression for the thermopower,

\begin{equation}
S=%
{\displaystyle {k_B \over e}}%
\left[ 
{\displaystyle {\epsilon _0 \over k_BT}}%
-\ln \left( \frac 54\right) -\ln \left( \frac{c(1-c)}{(1-2c)^2}\right)
\right]  \label{eqn2}
\end{equation}

\noindent where $\epsilon _0$ is the energy difference between identical
lattice distortions with and without the hole; the term $-(k_B/e)\ln
(5/4)=-19\;\mu V/K$ is associated with the spin entropy appropriate for a
spin-3/2 hole moving in a spin-2 background; and the last contribution is
the mixing entropy term in the case of correlated hopping with weak
near-neighbor repulsion \cite{chaikin}. At the nominal doping level $c=x=1/3$%
, this last term contributes $-60\;\mu V/K$; without the repulsive
interaction, the mixing term contributes $+60\;\mu V/K$ at the same hole
concentration.

$La_{2/3}Ca_{1/3}MnO_3$ films were deposited by laser ablation on $LaAlO_3$
substrates to a thickness of $0.6$ $\mu m$ and heat treated as described in
Ref. \cite{Jaime}. One of the films, annealed at $1000$ $^{\circ }C$ for $48$
hours in flowing oxygen, was mounted at the end of a stainless steel rod and
provided with current and voltage leads, and type-E thermocouples connected
in a differential mode. A nonmagnetically wrapped, temperature calibrated $%
12 $ $\mu m$ Pt wire, placed at one end of the sample, served both as a
heater to establish a thermal gradient across the sample, and as a
thermometer. The rod was placed in a $7$T Quantum Design {\em SQUID\ }%
magnetometer either with or without an oven option provided by the
manufacturer. In this way we were able to apply magnetic fields up to $7$T
and to vary the temperature in the range $4$ $K<T<500$ $K$. Following the
transport experiments, the magnetization ($M$) {\em vs.} temperature and
applied field was measured up to $380$ $K$ using conventional methods.

Fig. 2 displays $\ln (\rho /T)$ and $S$ versus $1000/T$ in a plot where it
is easy to see that both follow a thermal activated behavior with different
activation energies, $E_\rho \equiv $ $\epsilon _0+W_H-J=112${\rm \ }$meV$
and $E_s\equiv \epsilon _0=10$ $meV$ respectively. The $T\rightarrow \infty $
limit of the thermopower is $\simeq -30\;\mu V/K;$ Eq. (2) suggests that $%
c=0.29$ rather than \thinspace $0.33$ which could result from less than
nominal $Ca$ doping or, what is more likely, oxygen deficiency. There has
been considerable discussion in the literature concerning deviations of the
high-temperature extrapolation from the Heikes value, expected to be $%
+40\;\mu V/K$ at these concentrations.\cite{chaikin} However, in the
presence of hole-hole interactions, that value is approached only when $k_BT$
is large enough to overcome polaron repulsion, apparently beyond the
accessible experimental range. No $\rho \propto \exp (T^{-1/4})$ regime,
expected in the polaron picture for $T\leq \theta _D/2$, {\em i.e.} $T\sim
150-250$ $K$, was observed in the present study as the transition into a
metallic state at $T_C=238$ $K$ occurs above $\theta _D/2$. The activation
energies obtained at temperatures $T$ $>T_C$ give $W_H-J=102$ $meV$.
Assuming, as usual, that $J\ll W_H$ we note that the condition $%
T<W_H/k_B\approx 1190$ $K$ is satisfied. Use of the non-adiabatic
expressions results in activation energies $10\%$ larger, with similar fit
quality; we cannot, therefore, distinguish experimentally between the two
regimes.

That both $\ln (\rho /T)$ and $S$ deviate from linear behavior at $T\simeq
290$ $K$ in zero field provides evidence of identical microscopic
mechanisms, {\em i.e.} the onset of a long range order. From the
experimental values, the polaron formation energy is $E_p=2W_H\simeq 204$ $%
meV$ . Combining that value with expressions for $E_p$ in Ref. \cite{Moliton}%
, we obtain the relation $m^{*}/m_e=Ar_p^{-2}$ between the effective mass of
carriers $m^{*}$ and the polaron radius, with $A=1.9$ $nm^2$. This gives,
for example, a value $m^{*}/m\approx 3$ when $r_p$ is of the order of two
cell parameters. This value, relatively small for a localized electron in a
polaronic system, may reflect the significant role magnetic interactions
play in the self-localization process. The mass enhancement is a measure of
the {\em e-ph} coupling which, in Emin's model \cite{Emin}, is
counterbalanced by spin disorder, thereby causing polaron collapse.
Presumably, it is the delicate balance between {\em e-ph }coupling and
spin-disorder that causes $T_C$ to depend sensitively on doping and other
factors.

Because we find $E_\rho \neq E_S$, we can exclude the case of a Mott {\em MP}%
. In order to explore the possibility of a Kasuya {\em MP} we performed
resistivity and thermopower experiments under applied magnetic fields. The
results are displayed in Figs. 3({\em a}) and ({\em b}). In the temperature
range where the activation energy can be defined, there is a weak field
dependence of $E_\rho $ and $E_S$. Within experimental resolution, changes
in activation energies are different but of the same order of magnitude. An
estimation of average experimental values is $\triangle W_H/\triangle H=$ $%
-0.29$ $meV/T$ or $0.28$ $\%/T$ and $\triangle \epsilon _0/\triangle H=$ $%
-0.14$ $meV/T$ or $-1.4$ $\%/T$. While $\epsilon _0$ reflects changes in the
Fermi energy (see Fig. 1) that can be related to the reported
magnetostriction\cite{Ibarra} of {\em CMR} materials, changes in $W_H$ imply
an increase of the radius of the {\em HP} with field and consequently some
magnetic character of the quasiparticle. This may indicate that thermal
entropy limits the magnetic polarization of {\em HP} except in the proximity
of $T_C$, as predicted in Emin's model for polarons in ferromagnetic
systems. This small field dependence contrasts sharply with the effect of
external pressure,\cite{Khazeni} where a strong reduction of $E_\rho $ with
pressure was reported, perhaps reflecting the pressure dependence of the
transfer integral $J$ in Eq. (1).

In Fig. 4 we show the magnetization {\em vs.} temperature for several
magnetic fields in a similar temperature range. The data can be scaled as $%
M/(T-Tc)^\beta $ {\em vs.} $H/(T-Tc)^{\beta \delta }$ with Heisenberg-like
exponents, $\beta =0.38$, $\beta \delta \approx 1.8.$ There is considerable
similarity between the magnetization and the transport data in the vicinity
of $T_C$. We show this as an inset in Fig. 4, where the temperatures of
maximum slope in each quantity are plotted. At higher temperatures, the
resistance is much less field dependent than is the magnetization,
indicating that the exponential relation between $\rho $ and $M$ observed%
\cite{vonMol2} in {\em CMR} ceases at some point, marking perhaps a
crossover between regimes where the spin-polaron and the {\em HP} aspects
dominate.

Finally, as can be seen in Fig. 3({\em b}), the thermopower increases with
magnetic field at low temperatures. In order to explore the correlation
between $S$ and $\rho $ in the {\em FM}/metallic state we measured both as a
function of the applied field at a constant temperature $T=200$ $K$, as
shown in Fig. 5. Indeed, both properties change with applied field in the
proximity of $T_C$, even in the metallic state. Below $T_C$, the absolute
value of $S$ increases with field rather than decreasing sharply as observed
above $T_C$. We find that $S(H)\sim 1/\rho (H)$, a result that follows
fairly directly from the Nordheim-Gorter rule. If there are two sources of
resistance, and they add in series, then the thermopower contributions from
each process combine as $S=(\rho _1S_1+\rho _2S_2)/(\rho _1+\rho
_2)=S_1+\rho _2(S_2-S_1)/\rho $. \cite{Blatt} If we assume that only $\rho
_1 $ is field dependent, the observed result follows. A likely explanation
is that $\rho _1$ and $S_1$ are consequences of spin-disorder scattering,
which is reduced by an applied magnetic field.

In conclusion, we have presented experimental results showing strong
evidence for polaron transport in well annealed films of $%
La_{2/3}Ca_{1/3}MnO_3$ at temperatures double the transition temperature $%
T_C=238$ $K,$ significantly extending our preliminary results in partially
annealed samples. Our data allow us to rule out both Kasuya and Mott {\em MP}%
-dominated transport and the formation energy for {\em HP} was found to be $%
Ep=204$ $meV$. Although $\rho \sim \exp (-cM)$ behavior is not observed
above $T_C$, indications of a magnetic aspect to the polarons, from the
field dependence of their characteristic energies and relatively small mass,
suggests that they have both lattice and magnetic character. Even in the 
{\em FM}/metallic state just below $T_C,$ there remain significant
indications of spin scattering. Together these conclusions support an
intrinsic mechanism for the {\em CMR} effect, rather than extrinsic effects
due to granularity or other imperfections. Our results indicate that
polaronic collapse, such as treated by Emin, is close to the correct picture
for conduction in $LaCaMnO$ system. However, unlike the situation in $EuO$,
the collapse of large polarons in the ferromagnetic state reduces the
effective exchange coupling via the double exchange mechanism, causing a
``bootstrap'' destruction of ferromagnetism and a metal-insulator
transition. Recent theoretical models\cite{Roder} that combine
Jahn-Teller-driven lattice effects with the double exchange picture seem to
contain the necessary physics.

This work was prepared while MBS was a visitor at Tsukuba University with
the support of Mombusho. We would like to acknowledge Prof. M. Weissman for
useful discussions and the assistance of Ray Strange on high-temperature
aspects of this work. This research was supported in part by DOE Grant No.
DEFG0291ER45439 through the Illinois Materials Research Laboratory.\ 

\newpage\ \ \

\newpage\ \ \ 

\begin{figure}[tbp]
\caption{Band diagram for an electron-phonon coupled system with intrinsic
generation of carriers. $W_H$ is the energy required to jump in a given
direction, $\epsilon _o$ is the energy required to generate intrinsic
carriers, and $E_p$ is the binding energy of a polaron.}
\end{figure}
\begin{figure}[tbp]
\caption{The resistivity in the adiabatic regime and thermopower {\em vs.}
1000/temperature. Lines are fits in the high temperature regime, $E_\rho
\equiv \epsilon _0+W_H$ and $E_S\equiv \epsilon _0$ are the slopes.}
\end{figure}
\begin{figure}[tbp]
\caption{({\em a}) The resistivity in the adiabatic regime and ({\em b}) the
thermopower {\em vs. }the inverse of the temperature for different magnetic
fields.}
\end{figure}
\begin{figure}[tbp]
\caption{The magnetization {\em vs. }temperature for different magnetic
fields, after substracting a diamagnetic background. Inset: Transition
temperatures determined with the resistivity($T_\rho $), thermopower ($T_S$)
and magnetization($T_M$).}
\end{figure}
\begin{figure}[tbp]
\caption{The resistivity and thermopower at $T=200$ $K$ {\em vs.} magnetic
field. The solid line is a fit of the form $\alpha +\beta /\rho (H)$ with $%
\alpha =0.35$ $\mu VK^{-1}$ and $\beta =2.75$ $\mu V$ $m\Omega $ $cm$ $%
K^{-1} $. The dashed line is a guide for the eye.}
\end{figure}

\end{document}